\documentclass[prb,preprint]{revtex4-1} 

\usepackage{amsmath}  
\usepackage{amsfonts} 
\usepackage{graphicx} 
\usepackage{xcolor}
\usepackage[normalem]{ulem}  
\usepackage{cancel}
\usepackage{subcaption}
\usepackage{xcolor}

\begin{document}

\title{Spreading viscous fluids on a horizontal surface: \\ project-based learning in fluid mechanics.}

\author{R. Bolaños-Jiménez}
\email{rbolanos@ujaen.es} 
\altaffiliation[permanent address: ]{Campus Las Lagunillas, S/N, 23071, Jaén, Spain} 
\author{P. L. Luque-Escamilla} 
\affiliation{Department of Mechanical and Mining Engineering,Universidad de Jaén, Spain}


\begin{abstract}
The spreading of a thin viscous fluid film on a horizontal surface is an interesting problem in fluid mechanics with many practical applications ranging from coating processes to biological systems and environmental flows. It can even be observed in everyday situations, such as syrup spreading on a pancake.
We present a project-based learning approach to this problem, in which engineering or physics undergraduates apply classroom knowledge to understand and solve it, using dimensional analysis, experiments, and theoretical modeling. First, a dimensional analysis is conducted to guide the design of the experiment suitable for an undergraduate laboratory or even at home. The problem is then simplified to obtain a mathematical model that accounts for the experimental results. 
Through this process, students are able to obtain a solution compatible with those published in fluid mechanics journals with minimal supervision from the instructor. This project not only develops important skills but also motivates students by showing that they have the ability to solve complex problems.

\end{abstract}

\maketitle 

\section{Introduction} 

Project-based learning in engineering and physics strengthens conceptual understanding by connecting theory to real-world application while developing problem-solving ability, creativity, adaptability, and practical skills essential for tackling complex future challenges. 
Such project-based classes have been proposed and discussed in previous works in the field of fluid mechanics. \cite{alvaro,milla,williams,saleta,maroto} In the following sections, we describe the implementation of a fluid mechanics project through which students develop key skills, including dimensional analysis, experimental techniques (from the design of the experiment itself to data analysis), and theoretical modeling.

The problem we address here is the spreading of a thin viscous fluid film on a horizontal surface, which has been extensively studied in the literature \cite{Yeckel1994,Wilson2012,Lu2020,Devos2025}, and has many of practical applications \cite{craster} (\textit{e.g.,} coating and paints, inkjet printing, food technology, surface engineering, or oil spills. Understanding the dynamics of spreading liquid films is a cornerstone in the study of interfacial fluid mechanics. The final goal of the project is to obtain an expression for the radius of the liquid film, $R(t)$, in a simple way by following successive steps. Classical studies have addressed the spreading of a droplet driven by surface tension over a pre-wetted substrate, revealing the critical roles of precursor films and regularization strategies to resolve the contact line singularity~\citep{Bonn}. When the spreading is driven not only by capillarity but also by a continuous injection of fluid, as in purification devices or membrane coating, the flow exhibits complex behaviors governed by the interplay between injection rate, film thickness, and apparent contact line motion. \citet{kiradjiev} analyzed such surface-tension- and injection-driven spreading in thin films, obtaining power-law scalings for both the maximum film height and the contact line location, and revealing transitions in spreading dynamics depending on the time-dependence of the source. In contrast, gravity-driven spreading of viscous fluids, as studied by using lubrication theory, produces self-similar solutions with distinct scaling laws depending on the source type (constant flux or finite volume) and geometry (2D or axisymmetric), and highlights how buoyancy balances viscous dissipation.

Here, we describe how to make this problem accessible to undergraduate engineering or physics students, so that they can recover, with minimal guidance, the time evolution of the radius of a viscous fluid droplet spreading on a plane.

\section{Statement of the problem}

\begin{figure}[t]
\centering  
\includegraphics[width=7cm]{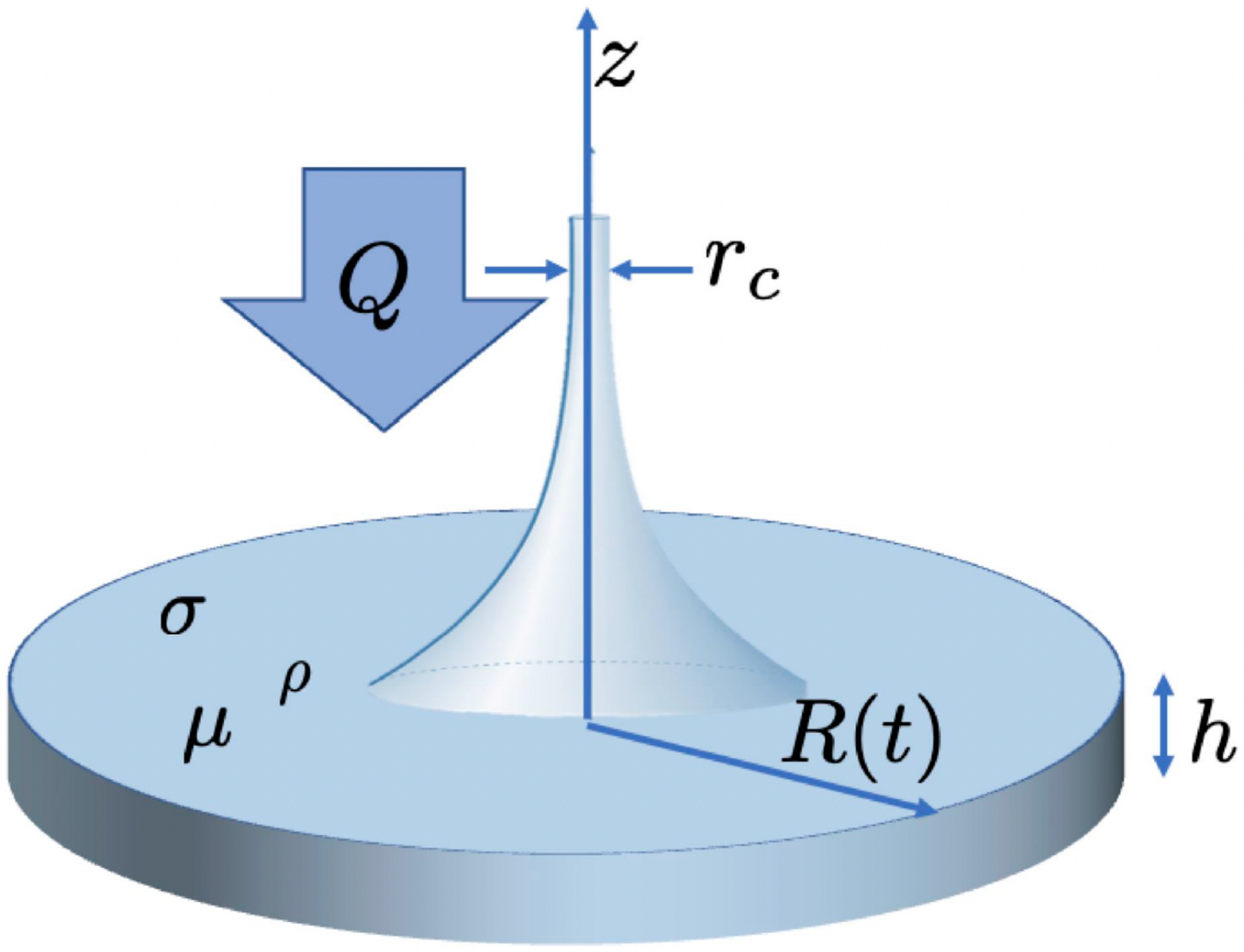} 
\caption{Sketch of the growing process of a fluid layer (density $\rho$, dynamic viscosity $\mu$ and surface tension coefficient $\sigma$) that falls at a constant volume rate $Q$ on a flat surface. The fluid spreads on the surface isotropically, creating a cylinder of radius $R(t)$ and thickness $h$.}\label{esquema1}
\end{figure}

The problem is sketched in Fig. \ref{esquema1}. A viscous, incompressible, and homogeneous liquid falls due to gravity perpendicular to a horizontal surface with a constant volume rate $Q$. Far enough from the surface, the jet has a radius $r_c$. 
Upon hitting the surface, the liquid, with density $\rho$, dynamic viscosity $\mu$, and surface tension $\sigma$, forms a thin circular layer of thickness $h$ and radius $R(t)$ on the horizontal surface. 

Our objective is to find an explicit expression for the spreading radius, $R$, as a function of time, $t$, and the governing fluid and flow parameters. To simplify the analysis to undergraduate students, we introduce the following assumptions.  
First, we will not focus on the initial moments of the phenomenon, nor on the regions close to the jet, and assume $R(t)\gg r_c$ and $R(t)\gg h$. 
Second, while in a real problem we would expect $h = h(t)$, here we will assume that $h$ is constant, as will be later verified. 
Third, to clearly observe the spreading process, we will concentrate on highly viscous fluids.
Fourth, we will assume that the fluid is Newtonian, which means that the viscosity $\mu$ is taken as constant and exhibits no elastic properties. 
Finally, we assume that the mean rugosity of the solid surface, $k$, is very small, $k\ll R(t)$. \footnote{Rugosity is a length that refers to the degree of roughness or structural complexity of a surface.}

The project is integrated into a Fluid Mechanics course and developed progressively alongside the theoretical content. Instructor intervention is intentionally limited, focusing on conceptual guidance rather than providing explicit solutions. 
The final project grade is based on the effort invested rather than on the outcome, encouraging creativity and enjoyment. Students will receive guidance and assistance throughout the process. However, the details of the implementation of the project may vary depending on the context and the preferences of the instructor. 
The solution corresponding to the three steps of this particular problem is detailed in the following sections. 

\section{Dimensional analysis}

As a first stage of the project, students perform a dimensional analysis of the problem, working in groups of three and submitting a short report after approximately one month, before the corresponding solution is discussed. The first step is to identify the physical quantities that control the dynamics of viscous liquid film spreading over a flat surface (see Fig.~\ref{esquema1}).  

Our objective is then to determine the mathematical 
relationship:
\begin{equation}
   R = f(t, \mu, \rho, \sigma, g, Q, r_c, h,k),\label{vars} 
\end{equation}
where $g$ is the gravitational acceleration, which drives the free fall of the fluid. 
If we adopt the canonical dimensional basis in Mechanics of mass, length, and time $\left\{ M,L,T \right\}$, the Buckingham theorem\footnote{The Buckingham $\pi$ theorem states that a dimensional relationship among $n$ variables can be rewritten as a relationship among $m=n-r$ dimensionless groups, where $r$ is the number of independent fundamental dimensions (e.g., mass $M$, length $L$, time $T$). \citep{buckingham}} tells us that we have $r = 3$ and, therefore, $m =7$ non-dimensional parameters. 
To obtain them, we can, for instance, follow the method proposed by Rayleigh.\citep{rayleigh}
Briefly, this involves expressing the dependent variable as a product of the independent 
variables raised to unknown exponents. By imposing dimensional homogeneity, these exponents are 
determined through a system of equations. Substituting them back allows the construction of 
dimensionless parameters from the original variables.
In this way, we get this first non-dimensional solution to our problem:

\begin{equation}
 \frac{R^3}{Q t}=\varphi\left(Fr, We, Re, \frac{k}{R}, \frac{h}{R},\frac{r_c}{R} \right),\label{analdim1}
\end{equation}
where $Fr = Q/\sqrt{g R^5}$, $We = \rho Q^2/(\sigma R^3)$ and $Re =\rho Q/(\mu R)$ can be identified with the well-known Froude, Weber and Reynolds numbers.

This first solution is too complex to be useful for experiment design, so we have to simplify it. 
To begin with, we can apply our  assumptions (see Section II), recalling that  $h \ll R$ and $R \gg r_c$. 
Therefore, we could neglect the influence of $h/R$ and $r_c/R$ in a first approximation. 
Moreover, for a highly viscous fluid,
 it is expected that the flow  will be laminar, with a small $Re$, 
 so that the inertial forces can be neglected in comparison with the viscous terms.

Recalling that the $Fr$, $We$, and $Re$ numbers are the ratios of inertial forces to gravity, surface tension and viscous forces, respectively, this assumption seems to imply that 
none of this parameters should be taken into account.
However, this is not true because neglecting all force-related parameters would result in losing all the dynamical information about the system. As stated before, the set of dimensionless quantities is not unique, so we can
rearrange these parameters to obtain a more convenient dimensional solution. In our case, this requires avoiding parameters depending on  inertia. 
This can be easily achieved by dividing. For instance, we can compute $Re/Fr^2$ to obtain $\rho g R^4/(\mu Q)$, which represents the ratio between theweight and the viscous force. 
Similarly, the product ${We^{-1}}{Fr^2} = \sigma /(\rho g R^2)$ represents the ratio between interfacial and gravitational forces and is the inverse of the well-known Bond number $Bo=\rho g R^2/\sigma$. We can then rewrite Eq. (\ref{analdim1}) as:
\begin{equation}
 \frac{R^3}{Q t}=\varphi\left(\frac{\rho g R^4}{\mu Q}, Bo, Fr, \frac{k}{R}, \frac{h}{R},\frac{r_c}{R} \right),
\end{equation}\label{eq1}
which, taking into account $k/R\ll1$, $h/R\ll1$, and $r_c/R\ll1$, can be finally written as:
\begin{equation}
\frac{R^3}{Q t}= \varphi\left(\frac{\rho g R^4}{\mu Q}\right),\label{analdim2}
\end{equation}
where $Fr$ and $Bo$ do not appear,  because gravity dominates  both interfacial forces and inertia  as we will now show.
In this problem, since $R$ is typically of the order of centimeters, the fluid density  similar to that of water,
and the surface tension   $\lesssim 10^{-1} \textrm{ N m}^{-1}$, then $Bo\gtrsim  10$, so that the interfacial forces are negligble compared to gravity. Moreover, 
a typical flow rate would be
 $Q \sim 1$ cm$^3/s$, so that $Fr \lesssim 10^{-1}$, which means that gravity dominates inertia. 
In addition, although the surface roughness has been neglected because $k/R \ll1$, it would not have influenced the fluid dynamics, as the flow is laminar. The influence of roughness at the contact line is only significant when surface tension dominates, which is not the case here.  
It is worth noting that all these  assumptions made during the dimensional analysis must be empirically confirmed. If experiments do not validate Eq. (\ref{analdim2}), then the simplifications need to be revised.

\section{Experiments}

Once the students have completed the dimensional analysis and the solutions have been shared, the next phase of the project, in which students design and perform experiments, becomes much easier. The goal of this second step is to determine the function $\varphi$ in Eq. (\ref{analdim2}). These experiments can be performed using common materials that are typically available at home. Using Eq. (\ref{analdim2}) as a guide, students obtain data pairs $\{R^3/(Qt),\rho g R^4/(\mu Q)\}$ by measuring the radius $R$ as a function of time $t$ for several values of the flow rate $Q$.

\begin{figure}[t]
\centering  
\includegraphics[scale=.52]{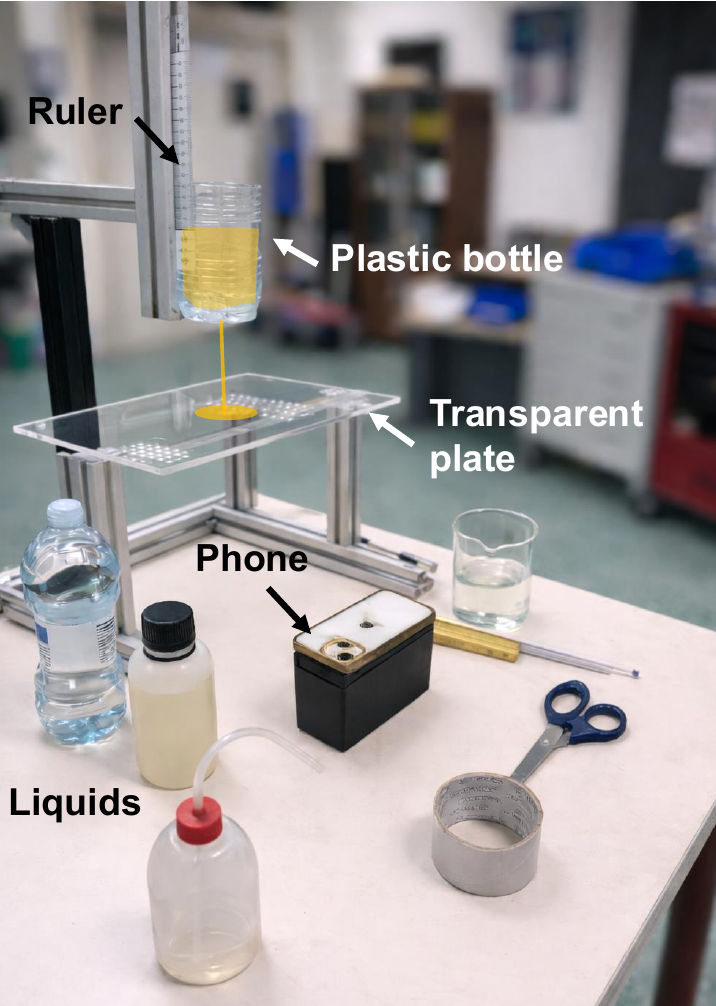} 
\caption{Experimental setup in our laboratory using easily accessible materials such as plastic bottles, mobile phone and tape.
}\label{setup} \label{setup}
\end{figure}

We propose that students use three different liquids: extra virgin olive oil, Fairy Professional\textcopyright\ dish soap, and a 30\%-70\% (w-w) solution of icing sugar in water. Their density and viscosity were measured 
at $T\simeq 20^\circ$ C using a \textit{Brookfield DV3TLVCJ0} rheometer and a \textit{Mettler Toledo Density2Go Densito} densimeter. These values were given to the students in advance (see Table \ref{tab:properties}). The values obtained for the olive oil~\citep{Abramovic, Bonnet, Peri}\textsuperscript{,}\footnotemark[25] and the icing sugar solution are in agreement with the results found in the literature, and those for the dish soap are consistent with the information provided by the manufacturer.\footnotetext[25]{For other sugar concentrations see \protect\url{https://lclane.net/text/sucrose.html}} 

\begin{table}[ht!]
    \centering
    \color{black} 
    \begin{tabular}{|c|c|c|c|}
        \hline
         & $\rho$ (kg/m$^3$) & $\mu$ (Pa$\cdot$s) & $\sigma$ (N/m) \\ \hline
        Virgin olive oil & 917 & 0.085 & 0.03 \\ \hline
        Dish soap & 1020 & 0.700 & 0.04 \\ \hline
        Water-icing sugar solution & 1347 & 0.489 & 0.06 \\ \hline
    \end{tabular}
    \caption{Physical properties of the liquids used in this work.}
    \label{tab:properties}
\end{table}

\begin{figure}[ht!]
\centering  
\includegraphics[width=8.5cm]{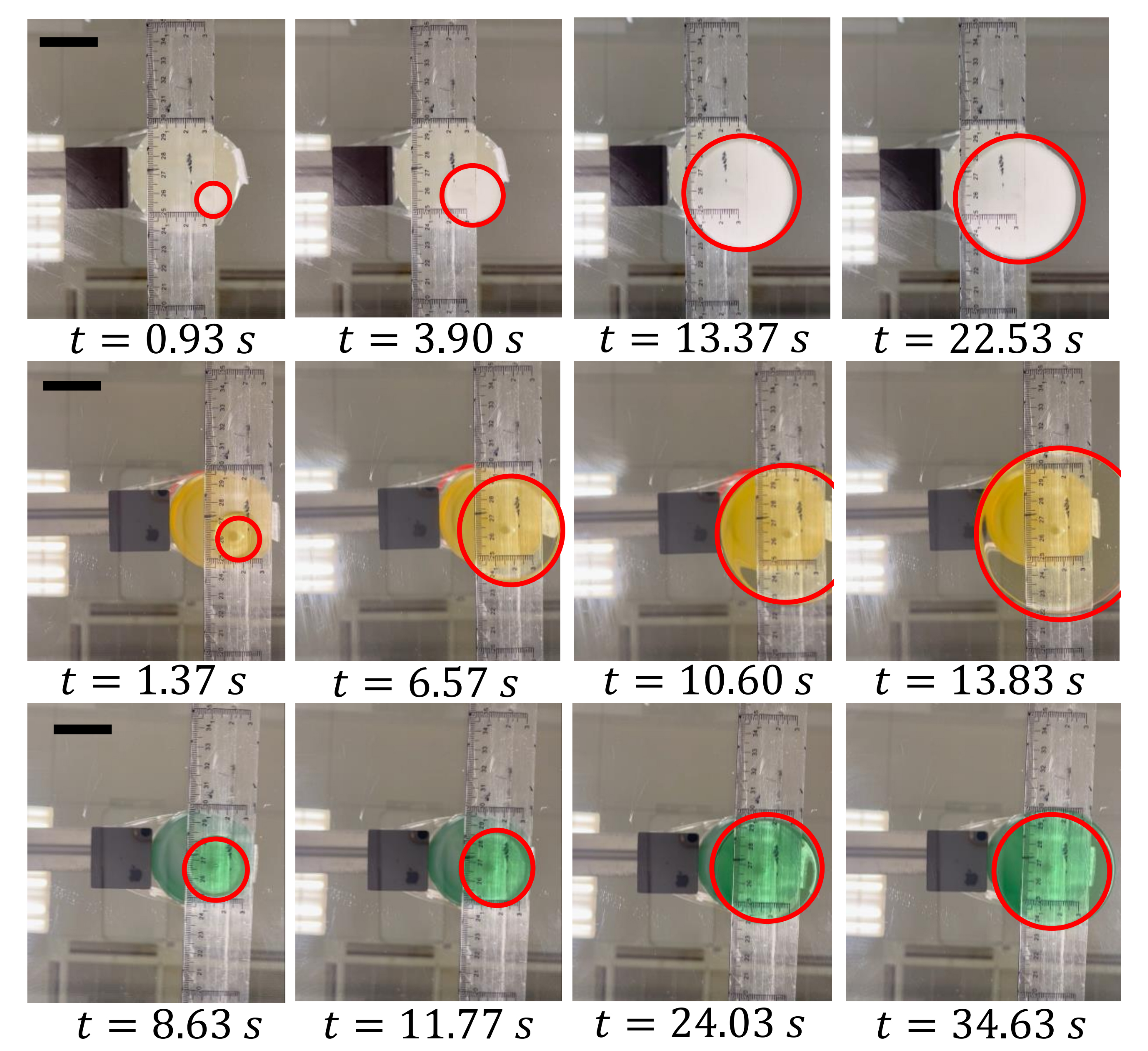}
\caption{Frames taken from the videos showing the expansion of the circular fluid plate in time. The top row corresponds to the sugar-water solution ($H_0$ = 1 cm), the middle to olive oil ($H_0$ =1.5 cm), and the bottom to the soap ($H_0$ =1.5 cm). Two circles are visible in each frame. The one with a fixed radius is the bottle, while the circle that gets larger in each panel, highlighted with a red circumference, is the liquid sheet of radius $R$($t$). The black line in left figures is a scale bar of 3 cm. }\label{fotos}
\end{figure}

The experimental set-up is deliberately simple (Fig. \ref{setup}). 
To generate a jet with a small and constant flow rate, we used a transparent plastic bottle with an opening of diameter $d$ at its base, which is assumed to be the diameter of the jet.
For the flow rate to remain constant, we must ensure that the liquid level changes very little during the experiment. For this, $d$ must be much smaller than the bottle's diameter. To measure the liquid level, students can place a ruler alongside the bottle, ensuring that both the bottle and the ruler are vertical. The radius of the circular liquid layer on the plate was measured by attaching a ruler to the transparent plate and recording videos with a mobile phone placed underneath. For the transparent plate, students are advised to use polymethyl methacrylate, polycarbonate or glass. These materials are smooth enough so that surface roughness does not influence the results. Fig. \ref{fotos} shows sample frames from our recordings. The flow rate of the jet must be small to ensure a low Reynolds number. 
We adjusted it by controlling the level of the liquid in the bottle, $H$, which ranged from 1 to 4 cm. The actual flow rate is $Q=C_d Q_i$, with $Q_i=(\pi d^2/4) \sqrt{2gH}$ being the ideal flow rate and $C_d = 0.141 Re^{1/2}$ the discharge coefficient\citep{kiljanski}. In our experiments, $0.05 \lesssim C_d \lesssim 0.4$ and $0.1 \lesssim Q \lesssim 180$ mL/min. 
The distance between the outlet orifice and the transparent plate should not be too large to prevent the jet from becoming unstable and breaking into droplets before impact\cite{frag}. We set it to approximately 5 cm.

For each liquid, we conducted multiple experiments by varying the initial liquid height, $H_0$, and consequently, the jet flow rate. During each experiment, we recorded two types of videos: one from underneath the transparent plate to measure the sheet radius as a function of time $R(t)$ (see Figs. \ref{setup} and \ref{fotos}), and one facing the fluid level in the bottle to observe changes in the height of the liquid 
$H$, with the camera field of view also including the surface where the jet impacts. 
The two videos were synchronized \textit{a posteriori}: the moment when the jet impacts the plate is visible in both recordings and is used as the time origin. 
An average height is used to compute the jet velocity, although in most cases the variations of $H$ are within the experimental uncertainty, since we aim to maintain a nearly constant flow rate.

To measure the radius of the liquid sheet at different times, $R(t)$, we used a free video editing program (\textit{ImageJ}). Specifically, we measured the sheet's diameter in pixels and, using the picture of the ruler, determined the conversion to length.  To measure time, we set the zero time as the moment when the jet impacts the surface. 
The mobile phone cameras recorded the videos at a fixed rate of 30 frames per second.  We started the measurements once the liquid layer formed a circle and stopped them when it lost its circular shape.

\begin{figure}[ht!]
\centering  
\includegraphics[width=8.5cm]{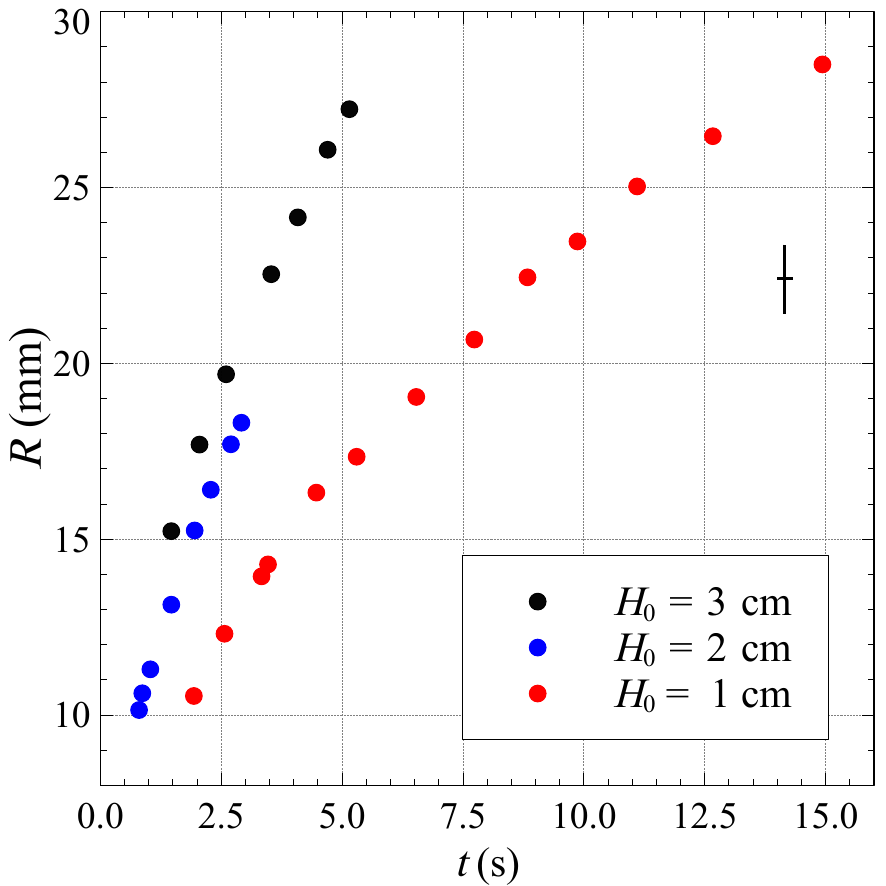}  
\caption{Radius of the liquid sheet as a function of time for a sugar--water mixture at different initial liquid heights, $H_0$. The error cross indicates representative uncertainties of 1~mm in $R$ and 0.033~s in $t$.
}\label{Rt}
\end{figure}

\begin{figure}[t!]
\centering  
\includegraphics[width=8.5cm]{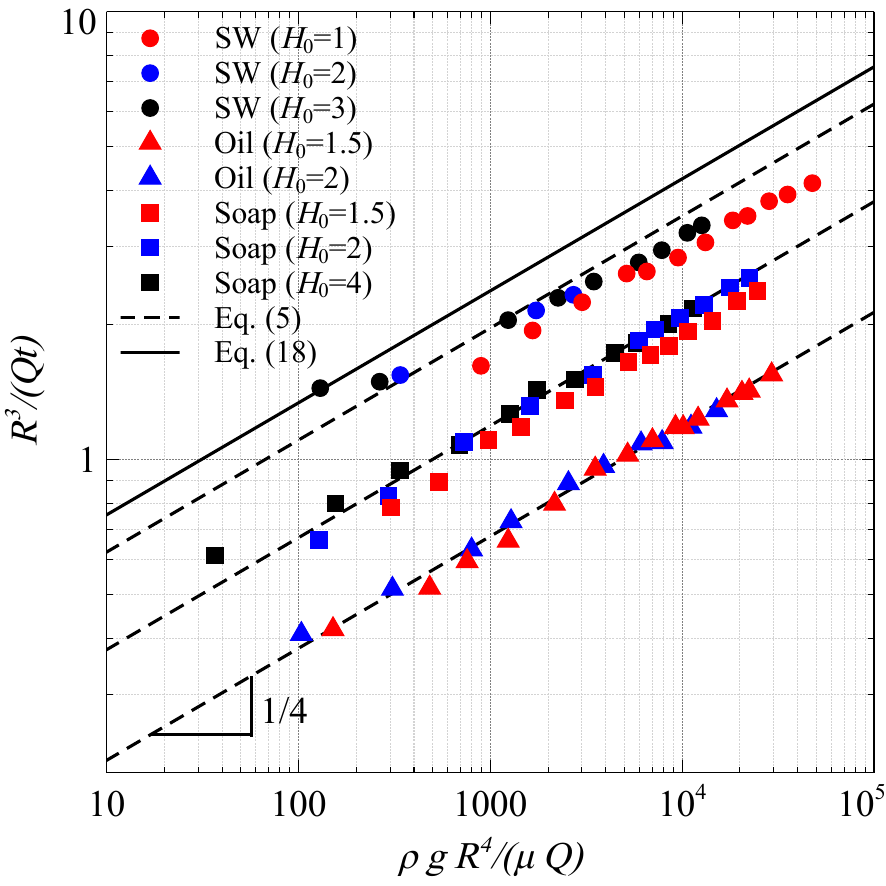}  
\caption{Experimental values of the dimensionless parameters for experiments on all three fluids. The equation obtained from dimensional analysis (Eq.~\ref{eq:solexp}) is represented by dashed lines, with $A=$0.35, 0.21, and 0.12 for sugar--water solutions, soap, and oil, respectively. Our theoretical model (Eq.~(\ref{def})) is also shown (solid line). Error bars are omitted for clarity; estimated uncertainties are discussed in the text.}\label{fig:final}
\end{figure}

Fig. \ref{Rt} shows  our measurements of the sheet radius as a function of time for the  sugar-water mixture. As expected, the higher the liquid level in the bottle (\textit{i.e.}, the greater the flow rate), the faster the sheet expands.   
According to the dimensional solution in Eq.~(\ref{analdim2}), the two parameters 
$R^3/(Qt)$ and $\rho g R^4/(\mu Q)$ are linked by a single functional relationship. 
Therefore, when the experimental data are plotted in this dimensionless form, all points 
are expected to collapse onto a single curve. 
However, when we plot the dimensionless variables, all our experimental values do not appear to follow a single curve. All data points for a given fluid collapse onto a single curve, showing that the dependence in flow rate is well taken into account. However, to each fluid corresponds a different curve, contrary to the dimensional analysis expectation (see Fig. \ref{fig:final}). 
Although the three curves appear to follow a common power-law behavior with identical exponent, the prefactor differs slightly, suggesting that some of the assumptions underlying the dimensional analysis in Section III may not be fully justified. In particular, while the analysis leading to Eq. (4) neglects several dimensionless groups—three purely geometric parameters and two fluid-dependent numbers, the Bond and Froude numbers—this neglect relies on the usual assumption that dimensionless groups taking extreme values have only a secondary influence and therefore do not appear explicitly in the final scaling law. 

In our experiments, however, the Froude number ranges from about $10^{-3}$ to $10^{-2}$, while the Bond number varies between $\sim 10^{-5}$ to $10^{-3}$, so both parameters indeed attain extreme values. The observed variation in the prefactor therefore suggests that, contrary to the standard expectation, at least one of these dimensionless groups may retain a residual influence. Among them, the Bond number is the most plausible candidate, since it depends explicitly on fluid properties. The fact that the fluid exhibiting the largest deviation is also the one with the highest surface tension further supports this interpretation. Nevertheless, the effect appears to be small: within experimental uncertainty the prefactor is compatible for all fluids, except possibly for olive oil. This issue will be revisited in the theoretical discussion section of the paper.

Overall, the data from different fluids and flow rates collapse onto a common scaling law, with only minor shifts in the prefactor. In particular, all experiments obey the relation:
\begin{equation}
     \frac{R^3}{Q t}=A\left(\frac{\rho g R^4}{\mu Q}\right)^{1/4}, \label{eq:solexp}
\end{equation}
with $A\sim 10^{-1}$ the  prefactor, wich seems to vary very slightly with $Bo$. In particular, $A$=0.35, 0.21 and 0.12 for sugar-water solutions, soap and oil, respectively (dashed lines in Fig.\ref{fig:final}). Finally, from Eq. (\ref{eq:solexp}) we obtain the dependence of the radius with time:
\begin{equation}
 R =A^{1/2} \left(\frac{\rho g Q^3}{ \mu }\right)^{1/8} t^{1/2}.\label{eq:def_exp}
 \end{equation}

For clarity, error bars are not displayed in Fig.~\ref{fig:final}. The dimensionless quantities shown in this figure are obtained from experimentally measured variables with associated uncertainties $\Delta R \simeq \Delta H \simeq 1$~mm and $\Delta t \simeq 0.033$~s, set by the spatial resolution and the frame rate of the recordings. The orifice diameter is measured using a calibrated drill bit, yielding $\Delta d \simeq 100~\mu$m. Using standard error propagation, the resulting relative uncertainty in the flow rate $Q$ is estimated to be of order 10\%, mainly due to the combined contributions of $C_d$, $d$, and $H$.
Propagation of these uncertainties leads to relative errors ranging from approximately 20\% to 50\% in both axes of Fig.~\ref{fig:final}, depending primarily on the value of $R$. The dominant contribution in both dimensionless variables originates from the determination of the spreading radius $R$, which enters with a high power. Consequently, uncertainties are largest at early times, when $R$ is smallest, and decrease as the sheet expands. For typical experimental conditions, relative uncertainties of about 20--30\% are obtained for $R^3/(Qt)$, while values up to 40--50\% may occur for $\rho g R^4/(\mu Q)$ at early stages of the spreading.
Despite these uncertainties, the experimental scatter remains sufficiently small to clearly identify the scaling behavior observed in Fig.~\ref{fig:final}, and in particular the characteristic $1/4$ slope in the log--log representation.

\section{Theoretical model}

Our goal is to obtain an approximate model for the growth rate $R(t)$ of the fluid film.
As stated in Section II, the fluid is assumed to be incompressible, homogeneous, laminar and Newtonian.
Let us consider a fixed and rigid control volume\footnote{A control volume is a fixed or moving region in space, bounded by a control surface, through which fluid may flow, and within which the fundamental conservation laws (mass, momentum, and energy) are applied.} as in Fig. \ref{esquema1}, which contains the fluid at time t, when the circular layer is assumed to be well-formed. We use a cylindrical reference frame with radial and axial unit vectors  $\hat{\rho}$ and $\hat{k}$ is used. The fluid enters the control volume through the upper circular surface $S_i = \pi r_c^2$, and exits through the lateral cylindrical surface, $S_o = 2\pi R h$. The volume of fluid inside the control volume is always constant at any time. To make the problem more manageable for students, we introduce reasonable simplifications. For instance, we assume $r_c/R\ll 1$ and $h/R\ll 1$ at all times, as we are interested in the steady behavior of the spreading. Since the fluid film is very thin, the lubrication approximation for laminar flow in shallow layers allows us to assume that the velocity profile is approximately linear in the vertical coordinate:
\begin{equation}
    \vec{v}=v_r(z)\hat{\rho}= \frac{ Q}{\pi R h^2}z\hat{\rho}.\label{perfilv}
\end{equation}
Moreover, as in the dimensional analysis, we assume a constant film thickness $h$, although, in reality, it is expected to decrease radially along the film. 
This assumption may reduce accuracy but greatly simplifies the problem.
Moreover, recall that inertia forces are expected to be negligible compared to gravity, resulting in $Fr \ll 1$.

\begin{figure}[ht]
\centering  
\includegraphics[width=8.5cm]{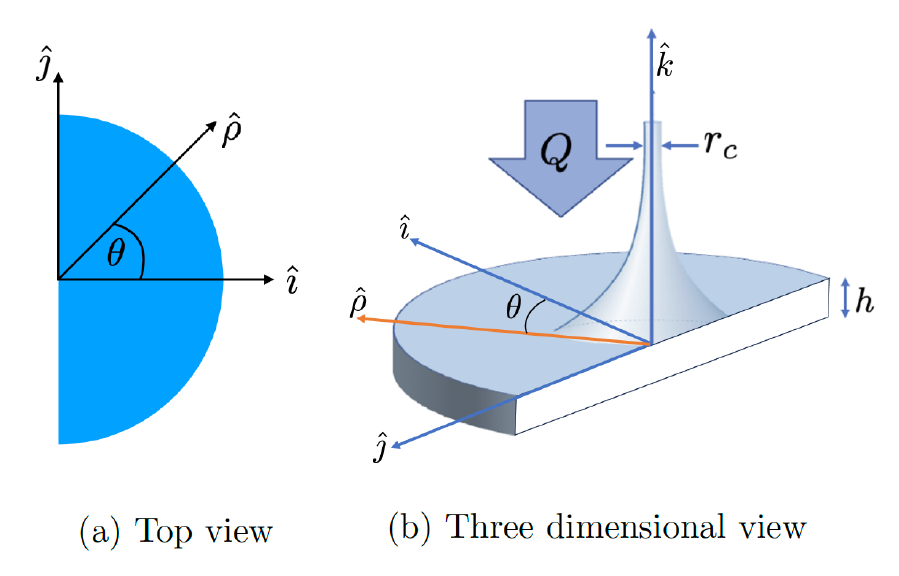}  
\caption{Schematic diagram of the control volume (and control surfaces) used in the conservation of momentum. }
\label{esq1}
\end{figure}

Conservation of mass \citep{white} then gives $Q = 2 \pi R h \dot{R}$, with $\dot{R}$ the spreading velocity. We can integrate this expression between $t=0$ and $t$ to obtain:
\begin{equation}
 R = \sqrt{\frac{Q}{\pi h}}t^{1/2}.\label{continuidad}
\end{equation}
Then, to derive the dependence of $Q$ on $h$, we use the conservation of momentum along the radial coordinate. 
However, due to symmetry, the same control volume cannot be used.
Therefore, we use a control volume consisting of half of the original volume, obtained by cutting it longitudinally, with the azimuthal angle $\theta$ ranging from 0 to $\pi$ (see Fig. \ref{esq1}).
In the steady state, the forces acting on the fluid (due to pressure $p$, viscous stress tensor $\boldsymbol{\tau}_v$ and gravity) are balanced by the convective flux of momentum. Because we are not interested in the axial direction, we apply the momentum conservation equation, that is, the Navier-Stokes equation in its integral form \citep{white} over the control surface (CS) of the control volume in the horizontal direction:
\begin{equation}
 \int_{CS} \rho \vec{v} (\vec{v}_r \cdot d\vec{S}) = -\int_{CS} p d\vec{S}+\int_{CS} \boldsymbol{\tau}_v d\vec{S}\label{ecmto}
\end{equation}
Let us calculate each integral separately. The left-hand side represents the flux momentum in the horizontal direction. Since the output flow passes through d$S_o = R d\theta dz$ we have:
\begin{equation}
\begin{split}
    \int_{CS} \rho \vec{v} (\vec{v}_r \cdot d\vec{S}) &=\int_{S_s} \rho v_r(z)^2 \hat{\rho} dS_o \\&=\frac{\rho Q^2}{\pi^2 R h^4} \int_{z=0}^{z=h} \int_{\theta = -\pi/2}^{\theta = \pi/2} z^2 \hat{\rho} d\theta dz
\end{split}
\end{equation}
As $\hat{\rho} = \cos\theta \hat{\imath} +\sin\theta\hat{\jmath}$, with i and j the usual unit vectors in the horizontal direction, we finally obtain:
\begin{equation}
 \int_{CS} \rho \vec{v} (\vec{v}_r \cdot d\vec{S}) = 
  \frac{2}{3}\frac{\rho Q^2}{ \pi^2 R h} \hat{\imath}.
\end{equation}

The pressure force in the horizontal direction can be  obtained by applying the mean value theorem, with $p_m = \rho g h/2$ representing the mean static pressure on the output surface:
\begin{equation}
    -\int_{S_s}p d\vec{S} = -p_{m} \int_{S_o} dS_o \hat{\rho} =  R h^2 \rho g (-\hat{\imath}).
 \end{equation}
Finally, the net viscous force acting on the horizontal plate can be calculated as the product of the viscous stress tensor $\bm{\tau}_v$ and the plate surface d$\vec{S}_h = \frac{R^2}{2}d\theta \hat{k}$. Since we are dealing with Newtonian fluids in laminar flow, the form of the viscous stress tensor is well-known. Using the symmetry of the system,  the only nonzero velocity is $v_r(z)$. Thus, we obtain:
 \begin{equation}
 \begin{split}
      \int_{CS} \bm{\tau}_v d\vec{S} &= \int_{S_h} \mu \frac{\partial v_r}{\partial z} dS_h =\hat{\rho}\frac{\mu Q R}{2 \pi h^2 } \int_{-\pi/2}^{\pi/2}d\theta \hat{\rho} \\ &= \frac{\mu Q R}{2 \pi h^2 }(2\hat{\imath} + 0 \hat{\jmath}) =  \frac{\mu Q R}{\pi h^2 }\hat{\imath}.
 \end{split}
 \end{equation}
 We finally express the momentum flux equation (Eq. \ref{ecmto}) as:
 \begin{equation}
      \frac{2}{3}\frac{\rho Q^2}{ \pi^2 R h} \hat{\imath} =  - R h^2 \rho g \hat{\imath} + \frac{\mu Q R}{\pi h^2 }\hat{\imath},
 \end{equation}
which can be written in scalar form as:
\begin{equation}
    (\pi^2 R^2  \rho g) h^4 + (\frac{2}{3}\rho Q^2) h = \pi R^2 \mu Q,
 \end{equation}
which can be rearranged taking into account the expressions for the Froude number ($Fr = v_o/(g h)$) and the mean velocity ($v_o=Q/(2\pi R h)$) at the outflow:
\begin{equation}
    h^4 (1+ \frac{8}{3}Fr^2) = \frac{\mu Q}{\pi \rho g}.
\end{equation}
Considering that we are neglecting the effects of inertia, and thus $Fr\ll1$, we arrive at an approximate  equation relating $h$ and $Q$:
 \begin{equation}
     h = \left( \frac{ \mu Q}{\pi  \rho g}\right)^{1/4}.\label{cantmvto}
 \end{equation}
For $Q$ ranging from 0.3 to 3 cm$^3$/s (our typical experimental values), the $h$  derived from Eq. (\ref{cantmvto}) varies between 1.45 and 1.71 mm, validating our assumption that $h \ll R$, as $R$ is typically a few centimeters in our experiments. 
In fact, the thickness in Eq. (\ref{cantmvto}) 
closely matches the mean value of the exact theoretical  solution $h(r)$\citep{huppert} for $r > 0.35 R$, which is consistent with our assumption that $R\gg r_c$.

Using Eqs. (\ref{continuidad}) and (\ref{cantmvto}), we can derive a simplified expression for the time-dependent radius of the fluid for a thin layer spreading on a flat surface:
\begin{equation}
 R= \left(\frac{\rho g Q^3}{\pi^3 \mu }\right)^{1/8} t^{1/2}= 0.651 \left(\frac{\rho g Q^3}{\mu }\right)^{1/8} t^{1/2}.\label{def}
 \end{equation}

We can compare this final result with the exact theoretical expression derived by \citet{huppert}:

\begin{equation}
    R =0.623 \left(\frac{\rho g Q^3}{\mu}\right)^{1/8} t^{1/2}.\label{solteoR}
\end{equation}
 
The close agreement obtained using our simple model based on a constant $h$ is noteworthy, suggesting that this assumption provides a robust and effective approximation for the problem at hand.

Moreover, 
Eq. (\ref{def}) is in agreement with the experimental solution in Eq. (\ref{eq:def_exp}), provided that $A=\pi^{-3/4}=0.42$.  
This solution is plotted as a solid line in Fig. \ref{fig:final}. 
Note that, although the theory correctly predicts the $t^{1/4}$
 scaling, it tends to overestimate the radius of the liquid layer at a given time.
 These differences may be accounted for by experimental uncertainties together with the weak dependence on the Bond number identified in Section IV, except possibly for the olive oil. 
Although the simplifying assumptions of the analytical model appear reasonable, this last empirical discrepancy must be explained. Therefore, it is worth revisiting these assumptions to verify their validity.
The Bond number $Bo \gtrsim 10$ and the Froude number $Fr \lesssim 10^{-1}$ in all our experiments,  supporting the assumption that the  dynamics is dominated by viscous and gravity forces. Surface roughness effects may appear both at the interface between the plate and the fluid and along the contact line at the front of the film. The former case is relevant if $Re > 1$, which is not our case, while the latter is significant only when surface tension plays a dominant role, \textit{i.e.,} $Bo\ll1$, which does not occur in our experiments. 
As other experimental results for the axisymmetric spreading of viscous liquids into air  strong supports  the theoretical solution in Eq. (\ref{solteoR}) (see the work of Huppert\citep{huppert} and references therein), which is very close to our approximate solution (Eq. \ref{def}), we suspect that olive oil exhibits a peculiarity that causes it to deviate from the dynamics described by  these equations. 
To investigate this, the  dynamic properties of the three tested fluids were compared. The Bond numbers are similar for all three fluids, suggesting that  the  surface tension has little influence. However,  $Re$ for olive oil exceeds by an order of magnitude the values found for the other tested fluids, particularly during the initial stages of the evolution  ($Re \simeq 3 \text{ \textit{vs}. } Re \simeq 0.1$). This, combined with the observation that $Fr$ is also lower for oil than that for soap or the water-sugar mixture ($Fr \simeq 0.1 \text{ \textit{vs}. } Fr \simeq 0.03$), indicates that inertial effects may play a role in those early stages. 
Therefore, despite being deliberately simplified to remain accessible to students, our theoretical analysis effectively captures the relevant physics and shows a remarkable agreement with experiments across the tested fluids.

\section{Conclusions}

In this paper, we present an example of project-based learning that tackles a problem of importance for industrial applications, such as fluid spillage on a horizontal surface. The activity follows the three steps typical  in engineering or physics practice. First, dimensional analysis is performed to gain a deeper understanding of the problem and to rationally simplify its complexity. Next, experiments generates data which, upon analysis, reveals an empirical relationship among the dimensionless parameters identified in the previous step. Finally, a theoretical analysis culminates in the development of a model that explains the experimental data.

In this  project students can follow every step. The theoretical model is  simplified but manages to capture the underlying physics at a level suitable for undergraduate students. Additionally, they have freedom in each step, from choosing variables for dimensional analysis to designing experiments, always under the guidance of the teacher. 
Students generally find the problem attractive and motivating, and they engage actively with the task. While the project is challenging, most students are able to make significant progress, although some require guidance from the instructor to complete the measurements or refine their analysis. Overall, the experience is very positive, as the activity encourages independent problem-solving and reinforces the core concepts of dimensional analysis and conservation laws.

\begin{acknowledgments}
The authors sincerely thank Arturo Aragón Buitrago, our laboratory technician, for his unwavering support of the experimental work.
\end{acknowledgments}

\section*{Author Declarations}
\subsection*{Author Contributions}
All authors contributed equally to this work.
\subsection*{Conflict of Interest}
The authors have no conflicts to disclose.\\

\end{document}